\documentclass[preprint,fleqn]{elsarticle}

\usepackage{lineno,hyperref}
\usepackage{amssymb,amsmath}
\usepackage{color}
\usepackage[flushleft]{threeparttable}
\usepackage{url}

\journal{Journal of \LaTeX\ Templates}

\biboptions{sort&compress}

\newcommand{\pol}{\varphi_{\rm pol}}
\newcommand{\rchi}{\chi^2/{\rm dof}}
\newcommand{\AH}{{\it Hitomi}}

\begin{document}
\begin{frontmatter}

\title{Study of the Polarimetric Performance of a Si/CdTe Semiconductor Compton Camera for the $\AH$ Satellite}


\author[address1]{Junichiro Katsuta}
\author[address1]{Ikumi Edahiro}
\author[address2]{Shin Watanabe}
\author[address2,address3]{Hirokazu Odaka}
\author[address2]{Yusuke Uchida}
\author[address1]{Nagomi Uchida}
\author[address1]{Tsunefumi Mizuno\corref{CA}}
\cortext[CA]{Corresponding author}
\ead{mizuno@hep01.hepl.hiroshima-u.ac.jp}
\author[address1]{Yasushi Fukazawa}
\author[address2,address4]{Katsuhiro Hayashi}
\author[address1]{Sho Habata}
\author[address3.5]{Yuto Ichinohe}
\author[address1]{Takao Kitaguchi}
\author[address1]{Masanori Ohno}
\author[address2]{Masayuki Ohta}
\author[address1]{Hiromitsu Takahashi}
\author[address2]{Tadayuki Takahashi}
\author[address5]{Shin'ichiro Takeda}
\author[address4]{Hiroyasu Tajima}
\author[address6]{Takayuki Yuasa}
\author[address7]{Masayoshi Itou}
\author{and SGD team}

\address[address1]{Hiroshima University, 1-3-1 Kagamiyama, Higashi-Hiroshima, Hiroshima 739-8526, Japan}
\address[address2]{Institute of Space and Astronautical Science, Japan Aerospace Exploration Agency,
3-1-1 Yoshinodai Sagamihara, Kanagawa 252-5210, Japan}
\address[address3]{KIPAC, Stanford University, Stanford, CA 94305, USA}
\address[address3.5]{Tokyo Metropolitan University, 1-1 Minami-Osawa, Hachioji, Tokyo 192-0397, Japan}
\address[address4]{Nagoya University, Furo-cho, Chikusa-ku, Nagoya, Aichi 464-8601, Japan}
\address[address5]{OIST Okinawa Institute of Science and Technology Graduate University,
1919-1 Tancha, Onna, Kunigami District, Okinawa 904-0495, Japan}
\address[address6]{RIKEN, 2-1 Hirosawa, Wako, Saitama 351-0198, Japan}
\address[address7]{
Japan Synchrotron Radiation Research Institute (JASRI), SPring-8, 1-1-1 Kouto, Sayo-cho, Sayo-gun, Hyogo 679-5198, Japan
}





\begin{abstract}
Gamma-ray polarization offers a unique probes into the geometry
of the $\gamma$-ray emission process in celestial objects.
The Soft Gamma-ray Detector (SGD) onboard the X-ray observatory \textit{Hitomi}
is a Si/CdTe Compton camera and is expected to be an excellent polarimeter,
as well as a highly sensitive spectrometer due to its good angular coverage
and resolution for Compton scattering.
A beam test of the final-prototype for the SGD Compton camera was conducted 
to demonstrate its polarimetric capability and to verify and calibrate the
Monte Carlo simulation of the instrument. The modulation factor of the
SGD prototype camera, evaluated for the inner and outer parts of the CdTe sensors as absorbers,
was measured to be 
0.649--0.701 (inner part) and 0.637--0.653 (outer part) at 122.2~keV
and
0.610--0.651 (inner part) and 0.564--0.592 (outer part) at 194.5~keV
at varying polarization angles with respect to the detector.
This indicates that the relative systematic uncertainty of the modulation factor 
is as small as ${\sim}3\%$.

\end{abstract}

\end{frontmatter}

\section{Introduction}
\label{sec: intro}

Gamma-ray polarization provides a unique probe into the geometry of the $\gamma$-ray emission process
in celestial objects, such as the geometries of accretion disks around stellar-mass black holes
and the magnetic field structures of pulsar and pulsar wind nebula.
Detection of $\gamma$-ray polarization from cosmological sources also provides
a stringent test of the vacuum birefringence effect resulting from some quantum gravity models.
For example, INTEGRAL/IBIS measurements reported a change in the polarization angle
after the Crab flare indicating magnetic field reconnection\,\cite{Moran16}.
IKAROS/GAPs
and INTEGRAL/IBIS observations of $\gamma$-ray polarizations from
$\gamma$-ray bursts (GRBs) have also placed the most stringent limits by far
on the
vacuum birefringence effect to date\,\cite{Yonetoku11,Yonetoku12,Gotz13}.
However, the INTEGRAL/IBIS measurements of the Crab polarization required
observations of more than $10^{6}~{\rm s}$ because IBIS was designed as a coded mask instrument
and was not optimized as a Compton camera. 
The GAPS measurements of the GRB polarization were
marginal because of insufficient statistics due to the limited scattering angle coverage 
and the poor angular resolution
under the severe resource constraints. 
To advance studies of $\gamma$-ray polarization
from celestial objects, $\gamma$-ray instruments with higher polarization sensitivities are required.

The Soft Gamma-ray Detector (SGD) onboard the \textit{Hitomi} satellite,
the sixth Japanese X-ray observatory launched on  February 17, 2016\,
\cite{Takahashi12,Takahashi14},
is expected to provide 10 times better sensitivity
in 60--600~keV than past and current 
observatories\,
\cite{Takahashi04-SGD,Tajima10,Watanabe12spie,Fukazawa14}.
The SGD achieves high sensitivities by suppressing the background
using a combination of the BGO active shield and
a Compton camera.
Internal backgrounds that cannot be rejected by the active shield can be suppressed by requiring 
consistency between the incident direction of the $\gamma$ ray inferred by the Compton kinematics
and the field of view defined by the collimator.
The Compton camera consists of multiple layers of silicon (Si) sensors\,
\cite{Tajima02,Fukazawa05_si,Takeda07}
surrounded by multiple layers of high-quality cadmium telluride (CdTe) sensors\,
\cite{Takahashi01b,Takeda12,Watanabe14}
to achieve good energy resolution and angular resolution for Compton kinematics.
This design includes a compact camera with high detection efficiency 
enclosed in an active shield of a reasonable size that retains a sizable effective area.

The SGD also provides information on the polarization of the incident $\gamma$-rays because the
Compton-scattering differential cross section depends on the azimuthal scattering angle 
with respect to the incident polarization vector. The SGD is also an excellent polarimeter
because of its good angular resolution due to its highly segmented semiconductor sensors,
and good scattering angle coverage\,\cite{Tajima03,Takeda10}

To conduct reliable polarimetric measurements in orbit,
a precise characterization of the Compton camera 
for polarized $\gamma$ rays using a Monte Carlo (MC) simulation
with a detailed model of the instrument is crucial. Therefore,
we conducted a series of beam tests of prototypes for the SGD Compton camera using highly polarized
(${\ge}99\%$) $\gamma$ rays at the SPring-8\footnote{\url{http://www.spring8.or.jp}} synchrotron radiation facility 
to demonstrate
the polarimetric capability of the instrument, and to verify and calibrate the MC simulator.
As reported by \cite{Takeda10}, we evaluated the polarimetric performance of an
early prototype for the SGD Compton camera,
and verified that the observed azimuthal scattering angle distribution  
agrees well with the MC simulation. In November 2015, we conducted a beam test
using the final-prototype for the SGD Compton camera whose design is the same as that of the flight hardware.

In this study, we demonstrate the polarimetric capability of the SGD Compton camera
and verify and calibrate the MC simulation based
on the experimental data from the 2015 beam test.
In Section\,\ref{sec: method},
we briefly describe a method to derive polarization information using a Compton camera.
A detailed description of the SGD Compton camera is given in Section\,\ref{sec: camera}.
Section\,\ref{sec: setup} describes the setup and procedures for the beam test at SPring-8.
In Section\,\ref{sec: ana}, we provide the experimental results and comparisons
with the simulations.

\clearpage
\section{Method of Polarimetric Measurement}
\label{sec: method}

In this experiment, we adopted the standard method of polarimetric measurement 
for Compton polarimeters as described in Takeda et al. \cite{Takeda10}.
They derived the polarization information of the incident $\gamma$ rays from
a measured azimuth scattering angle ($\phi$) distribution based on the 
dependence of the Compton-scattering differential cross section on the azimuthal scattering angle
$\eta$ with respect to the electric vector of the incident $\gamma$ ray:
\begin{linenomath}
\begin{equation}
 \label{eq: cross}
  \frac{d\sigma}{d\Omega} = \frac{r^2_e}{2}
  \left(\frac{E'}{E_0}\right)^2\left(\frac{E'}{E_0} + \frac{E_0}{E'} - 2\sin^2{\theta}\cos^2{\eta}\right)~~,
  \end{equation}
\end{linenomath}
in which,
\begin{linenomath}
\begin{equation}
 \label{eq: e1}
  E' = \frac{E_0}{1 + (1 - \cos{\theta})E_0/m_e c^2}~~,
\end{equation}
\end{linenomath}
where $r_e$ is the classical electron radius, $m_e c^2$ is the electron rest mass,
$\theta$ is a scattering polar angle,
and $E_0$ and $E'$ are the energies of the incident and the scattered $\gamma$ rays respectively.
Below we briefly describe the analysis procedure
(see \citet{Takeda10} and references therein for details).

First, we measure the azimuthal angle distribution, $N_{\rm obs}(\phi)$,
using the Si/CdTe Compton camera.
Because the obtained distribution is affected by the detector response,
$N_{\rm obs}(\phi)$ needs to be divided by 
the azimuthal angle distribution for nonpolarized $\gamma$ rays, $N_{\rm iso}(\phi)$:
\begin{linenomath}
\begin{equation}
 \label{eq: n_cor_obs}
  N_{\rm cor}(\phi) = \frac{N_{\rm obs}(\phi)}{{N_{\rm iso}(\phi)} / ({\overline{N_{\rm iso}}})},
\end{equation}
\end{linenomath}
where $\overline{N_{\rm iso}}$ is the average per angle bin.

By referring to Equation~(1), we can see that the
corrected azimuthal angle distribution follows a sinusoidal curve:
\begin{linenomath}
\begin{equation}
 \label{eq: n_cor_mdl}
  N_{\rm cor}(\phi) = A(1 + Q\cos2(\phi - \phi_0 - \pi/{2})),
\end{equation}
\end{linenomath}
where $A$ is the normalization, $Q$ is the so-called modulation factor, and
$\phi_0$ is the direction of the polarization vector
with respect to the electric vector of the incident photon\,\cite{Lei97}.
%
By fitting the experimental data with this model formula, 
we can derive the polarimetric parameters, i.e.,
$\phi_{0}$ gives the polarization vector of the incident photon
and $Q$ is related to the polarization degree $\Pi$ such that
\begin{linenomath}
\begin{equation}
 \label{eq: pol}
  \Pi = \frac{Q}{Q_{100}},
\end{equation}
\end{linenomath}
where $Q_{100}$ (i.e., the analyzing power) is
the modulation factor for 100\% linearly polarized $\gamma$ rays.
The accuracy of the analyzing power affects the sensitivity
of the polarimetric measurement of the detector.
In Section\,\ref{sec: ana}, we evaluate the accuracy of $Q_{100}$ 
in the SGD Compton camera prototype
via a comparison between the experimental data and the simulated data.

\clearpage
\section{Compton Camera of SGD}
\label{sec: camera}

\subsection{Design}
\label{sec: design}
\begin{figure}[h] 
 \begin{center}
 \includegraphics[width=2.5in]{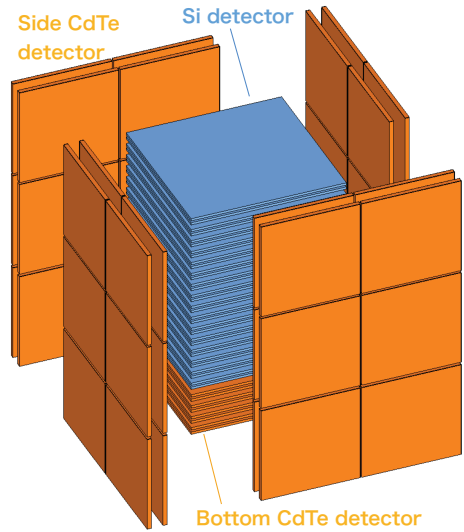}
\caption{
Conceptual image of the Compton camera.
Positions of the Si and CdTe detectors are shown.
\label{fig: cc_geo1}}
 \end{center}
\end{figure}

\begin{figure}[h] 
 \begin{center}
 \includegraphics[width=6in]{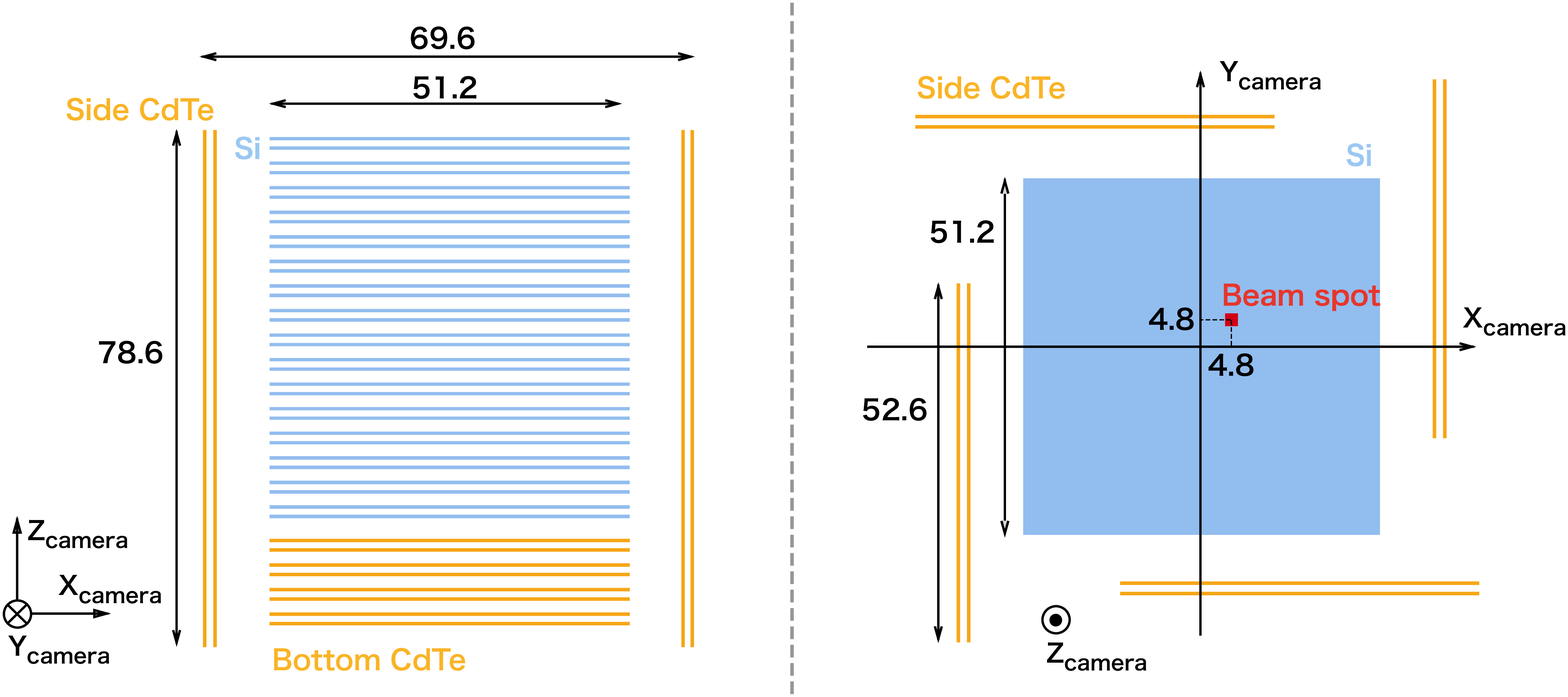}
\caption{
(Left) Side view of the Compton camera. The numbers are in millimeters.
(Right) Overhead view of the Compton camera.
The red square represents the incident beam spot (see Section\,\ref{sec: setup} for details).
\label{fig: cc_geo2}}
 \end{center}
\end{figure}

Here we briefly summarize the design of the Compton camera (see \cite{Watanabe14} for details).
The camera consists of top 32 layers of Si pixel sensors, 
bottom 8 layers, and 2 layers on each side of the CdTe pixel sensors, as shown in Figure\,\ref{fig: cc_geo1}.
The bottom and the side layers are composed of $2 \times 2$ and $2 \times 3$ CdTe sensors respectively.
The Si and the CdTe sensors have $16 \times 16$ pixels and $8 \times 8$ pixels, respectively,
with respective thicknesses of 0.6 and 0.75\,mm.
The pixel size is $3.2 \times 3.2$\,mm$^2$ for both the Si and CdTe sensors.
Figure\,\ref{fig: cc_geo2} displays the geometry of the camera with definitions of
instrumental (camera) $X$-, $Y$-, and $Z$-axes.

Each set of $8 \times 8$ pixels is read by a single ASIC (VATA450.3) mounted on 
a Front-End Card (FEC). 
The front-end electronics of the Compton camera consists of four groups of 52 FECs,
one ASIC Driver Board (ADB),
and one ASIC Control Board (ACB).
Eight or six ASICs are daisy chained for readout and control.
The ASICs are controlled by an FPGA on the ACB. 
These components are packed into a $12\times12\times12$ cm$^3$ aluminum box.
One of the daisy chained readouts did not work for the camera used in this experiment;
therefore, one fourth of the pixels in the eight Si layers cannot detect $\gamma$ rays.
However, this is not the case for the flight-model SGDs.

\clearpage
\subsection{Event Reconstruction}
\label{sec: rec}
When observing weak objects in orbit, we require that each SGD event involve Compton scattering in the detector
to apply Compton kinematics. The simplest case is that the event interacts twice in the camera:
once via Compton scattering in a Si sensor and once 
via photo-absorption in a CdTe sensor.
The time width to identify coincidence hits is ${\sim}10~{\mu s}$,
and typical low energy thresholds for hits in a Si sensor and a CdTe sensor are ${\sim}5$ and ${\sim}10~{\rm  keV}$, 
respectively\,\cite{Watanabe14}.
Once the locations and energies of the two interactions are measured,
the Compton kinematics allows us to calculate the direction of the incident photon
using the formula,
\begin{linenomath}
\begin{equation}
\cos{\theta} = 1+\frac{m_{e}c^{2}}{E_{1}+E'}-\frac{m_{e}c^{2}}{E'}~~,
\end{equation}
\end{linenomath}
where $E_{1}$ and $E'$ are the energy deposited in the Compton site and the photo-absorption site,
respectively, and are related to the incident photon energy such that $E_{0} = E_{1} + E'$ [see also Equation~(1)].
The deposit energies and positions are precisely measured
with  
energy resolutions ($\Delta E$) in full width at half maximum (FWHM) of ${\sim}2$ and ${\sim}3$~keV\footnote{
CdTe sensor has a low-energy tail in its spectrum and the energy resolution
was evaluated with the tail taken into account in experiments.
In simulation we modeled the spectrum with a gaussian for simplicity (see also Section~3.3).
}
at 122 keV for a Si sensor and a CdTe sensor, respectively, and
a position resolution of 3.2\,mm.
By requiring that the incident photon direction inferred from the Compton kinematics 
is consistent
with its presence within the field-of-view of the SGD, we can distinguish target signals from the background.
In addition, we apply a constraint on the topology and energetics of the detected events
to efficiently reject background signals while retaining a large effective area.
The details of the SGD background rejection can be found in\,\cite{Ichinohe16}.

\clearpage

\subsection{Monte Carlo Simulator}
\label{sec: simulation}
Unlike the monoenergetic pencil beam available at a synchrotron facility such as SPring-8,
celestial $\gamma$ rays have a continuous energy spectrum
and are irradiated on the entire detector surface area in orbit.
Therefore, Monte Carlo simulations are essential to construct the detector response.
We calculated the SGD instrumental response using
ComptonSoft\footnote{https://github.com/odakahirokazu/ComptonSoft/}\,\cite{Odaka10},
a Monte Carlo simulator that considers the detailed geometry of the Compton camera,
including the passive materials.
ComptonSoft includes Monte Carlo simulations and data analyses of Compton cameras,
which are optimized for multilayer semiconductor Compton cameras in particular.
It uses the Geant4 toolkit library \cite{Agostinelli03, Allison06} as the Monte Carlo engine,
and we adopted a low-energy electromagnetic physics option valid for polarized photons
(class \texttt{G4EmLivermorePolarizedPhysics} is used as the physics list constructor in Geant4). 
We carefully built a mass model of the SGD Compton camera
because a mass model is essential to produce the instrumental response accurately.
We included not only the active detectors but also the passive materials,
such as the supporting structures
and the electrical circuit substrates in the mass model.
Still there may be additional unaccounted for materials,
e.g., small amounts of heavy metals inside the readout 
electronics, that affect photon absorption. We discuss this issue further in Section\,\ref{sec: ana}.

We performed simulations with $(2\mbox{--}4) \times 10^{7}$ photons 
to obtain statistically sufficient references 
for each measured energy (122.2 and 194.5~keV; see Section~4). 
The properties of the incident photons---energy, direction, and the size of the beam---
were set to those in the experiment. The simulation code is capable of treating pixelated readout;
therefore, we eliminated channels with malfunctions and/or with noises that could decrease the sensitivity to the polarization. 
The spectral response of each readout channel was found to be uniform;
therefore, we assigned common energy-resolution parameters to all the readout channels of the same type of detector.
The energy resolutions 
(FWHM)
of the Si and the CdTe detectors are
\begin{linenomath}
\begin{gather}
\Delta E_\mathrm{Si} = 2.35 \times \sqrt{(0.80\;\mathrm{keV})^2+(0.005E)^2}, \\
\Delta E_\mathrm{CdTe} = 2.35 \times \sqrt{(0.82\;\mathrm{keV})^2+(0.0079E)^2},
\end{gather}
\end{linenomath}
where $E$ is the energy deposited in a pixel.
Even though it is known that a CdTe detector has a low-energy tail in its spectrum due to inefficient
charge transport (see also, e.g., \cite{Odaka10}), we can neglect this effect 
when examining the polarimetric response for this experiment 
because we applied a simple energy cut in the event selection,
as described in Section\,\ref{sec: ana}.
We therefore simulated the spectrum of both Si and CdTe sensors assuming a gaussian response.

\clearpage

\section{Setup and Procedure for the Experiment}
\label{sec: setup}
\begin{figure}[h] 
 \begin{center}
 \includegraphics[width=4in]{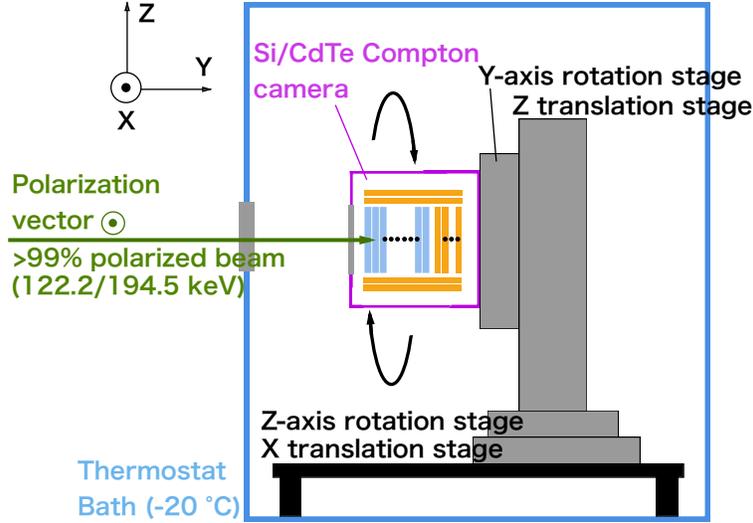}
\caption{
Schematic view of the experimental setup at Beamline BL08W.
\label{fig: setup1}}
 \end{center}
\end{figure}

The experiment was conducted at the High Energy Inelastic Scattering Beamline BL08W \cite{BL08W1,BL08W2,BL08W3}
of SPring-8 from November 12--14, 2015.
The experimental setup is illustrated in Figure\,\ref{fig: setup1},
with definitions of $X$-, $Y$-, and $Z$-axes of the experiment.
As shown in the figure, the camera was located in a thermostat bath,
which kept the temperature at 
$-20^{\circ}$C, the typical operational temperature of the SGD
onboard the \textit{Hitomi} satellite in orbit.
For the rotation and alignment of the camera,
we used motorized stages, as summarized in
Table\,\ref{tbl: stages}.

The incident beam was lineally polarized at ${\ge}99\%$ in the horizontal direction
and was irradiated perpendicular to the Si layers of the camera.
Although the SGD is expected to provide good sensitivity in 
60--600~keV\,\cite{Takahashi04-SGD,Tajima10,Watanabe12spie,Fukazawa14}, we
conducted the experiment using beam energy of 122.2 and then 194.5\,keV.
The beam energy of 122.2\,keV was selected because 
the peak of the effective area
of the SGD is near 120\,keV\,\cite{Watanabe14},
and the 122\,keV peak from the radiation source $^{57}$Co was used for the energy calibration.
We used the 194.5\,keV beam to examine the energy dependence of the 
polarimetric measurement capability of the Compton camera.
For each energy, the Compton camera was rotated along the beam axis using
$Y$-axis rotation stage
to evaluate the camera response for various directions of the polarization vector.
The experimentally set polarization angles ($\pol$) 
with respect to the camera's $X$-axis
were $0^{\circ}$, $22.5^{\circ}$, $45^{\circ}$, $90^{\circ}$, and $180^{\circ}$ (see Figures~2 and 4).
We primarily measured the data for $\pol = 0^{\circ}\mbox{--}90^{\circ}$
because the camera has a $90^{\circ}$-symmetric structure around the beam axis.
To check the consistency of the experimental setup,
we also obtained data for $\pol = 180^{\circ}$, which should be consistent with that of $\pol = 0^{\circ}$.
For each angle, we aligned the camera with the beam so that the incident beam ran through the same spot of the camera
and along the camera's $Z$-axis.
The aligned spot was set at a pixel whose center was located 
$4.8\,{\rm mm}$ from the center of the Si layers along the camera's $X$- and $Y$-axes 
(see also Figures\,\ref{fig: cc_geo2} and 4)
to avoid gaps between the bottom CdTe sensors under the Si layers
because each CdTe layer is composed of $2 \times 2$ sensors (see Section\,\ref{sec: design}).

The beam size was approximately 0.8\,mm wide and 1.4\,mm high.
The intensity of the beam was kept constant for each energy.
The event rate of the camera 
(either single hit event or multiple hits event)
was $\sim640$\,Hz; the background rate (environmental radio activities)
was $\sim 60$\,Hz so that the net event rate was 
$\sim580$\,Hz, giving the reconstructed event rate of ${\sim}20$ and ${\sim}30~{\rm Hz}$
for 122.2 and 194.5~keV beam, respectively (see also Section~5.1).
Even though the precise incident beam intensity was not measured, 
we could roughly estimate the intensity to be approximately $580\,{\rm Hz} / sp / (1-d) \sim 1200\,{\rm photons/s}$,
where $sp=0.92$ is the detector stopping power and $d=0.47$ is the dead-time ratio at 122.2\,keV.
The dead-time ratio was very high due to the high-intensity incident beam.\footnote{
Although the dead-time rate was high, we measured the live precisely (at the ${\le}2\%$ level) using a
``preudo trigger'' function\,\cite{Watanabe14}. Therefore the impact on the validation of the
instrument and simulator is small.}
The integration time for each polarization angle was approximately 80\,min.
The integration time was determined to obtain the modulation factor
with a statistical error of ${\le}1\%$.
To check whether the experiment was reproducible,
we measured the camera responses for $\pol = 0^{\circ}$ at the beginning and end of the experiment.
Both data sets were consistent within statistical errors at the energies of 122.2 and 194.5\,keV.

\begin{figure}[h] 
 \begin{center}
 \includegraphics[width=4in]{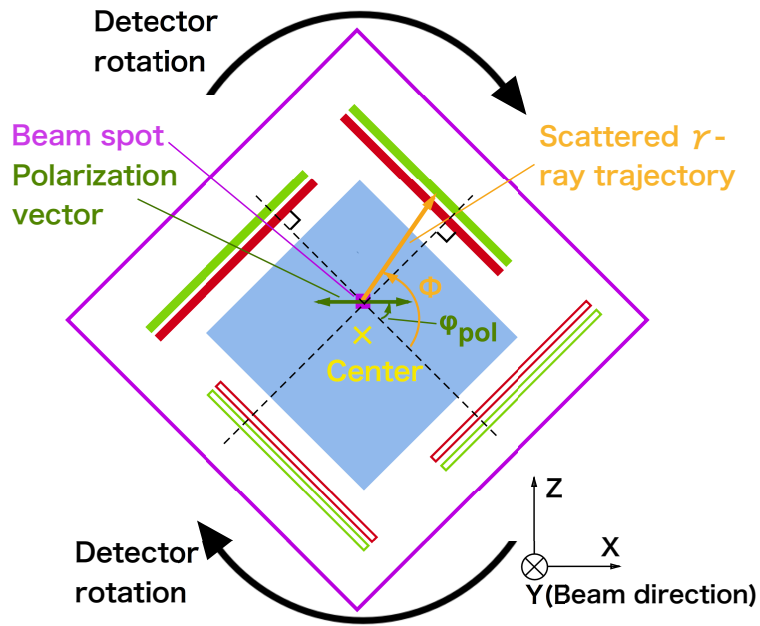}
\caption{
Top view of the Compton camera.
The definitions of the azimuthal scattering angle $\phi$ and the polarization angle $\varphi_{\rm pol}$,
measured with respect to the camera's $X$-axis, are shown (see also Figure~2).
The coordinate system of the experiment is the same as that in Figure\,\ref{fig: setup1}.
The inner and the outer four side CdTe layers are drawn as rectangles
in red and green, respectively.
The CdTe layers that are relatively closer to the incident beam spot are filled in 
with red or green,
while the other relatively-distant layers are filled in with white (see Section\,\ref{sec: ana_sim}).
\label{fig: setup2}}
 \end{center}
\end{figure}

\begin{table}[h]
 \begin{center}
 \begin{threeparttable}
  \caption{Stages used for the experiment} 
  \begin{tabular}{|l|l|r||r|} \hline
  Sort & Model number & Resolution\\ \hline
  X translation stage & XA16A-R1 & 1 ($\mu$m/step) \\
  Z translation stage & ZA16A-X1 & 1 ($\mu$m/step) \\
  Y-axis rotation stage & RA16A-WH & $2 \times 10^{-3}$ (deg/step) \\
  Z-axis rotation stage & RA10A-T01 & $5 \times 10^{-4}$ (deg/step) \\ \hline
  \end{tabular}
   \begin{tablenotes}
\small \item All stages were supplied by Kohzu Precision Co. Ltd. (Japan)
modified for use at a temperature of $-20^{\circ}$C.
\end{tablenotes}
  \label{tbl: stages}
 \end{threeparttable}
 \end{center}
\end{table}

\clearpage

\section{Analysis and Results}
\label{sec: ana}

In this section, we demonstrate and evaluate the polarimetric performance of the SGD Compton camera.
First, we produce the simulation data and the $N_{\rm cor}(\phi)$ distribution
and fit it with the model curve as described in Section\,\ref{sec: method},
to understand the polarimetric response of the SGD Compton camera.
Then, we compare the experimental data with the simulation data to tune the MC simulation and
to evaluate the systematic uncertainty of the SGD Compton camera.
Finally, we evaluate the sensitivity of the polarized signal from
a well-known celestial object, the Crab Nebula.

\subsection{Event Selection}
\label{sec: selection}
To measure the beam polarization,
we selected the reconstructed events (see also Section\,\ref{sec: rec}) 
that met the following conditions:
(i) The deposited energies in the Si sensor were 5--40~keV and 5--80~keV for the 122.2 and 194.5~keV beam,
respectively.
(ii) The measured total deposit energy was within 10~keV of the incident beam energy 
(122.2 or 194.5\,keV).
(iii) The incident $\gamma$ ray was scattered by the Si sensors and then absorbed by the side CdTe sensors.
(iv) The hit position in the Si sensor was at the pixel of the beam spot.
Even though including the bottom CdTe sensors increases the number of events,
we only used the side CdTe sensors as an absorber in this analysis for simplicity.
We note that including the events with hits in bottom CdTe sensors is an important next step for
fully validationg the instrument and simulator, and is also crucial to improve the statistics of the observational data.
The selected events were approximately 6\% and 4\% of all the detected events 
from the beam (either single hit event or multiple hits event)
at 122.2 and 194.5\,keV, respectively.

\clearpage

\subsection{Simulation Data}
\label{sec: ana_sim}
We created simulation data in which the incident $\gamma$ rays went along the $Y$-axis with
an energy of 122.2\,keV,
a physical size of $0.8 \times 1.4~{\rm mm^{2}}$,
and a polarization degree of 100\% with $\pol = 0^{\circ}$ with respect to the
camera's $X$-axis, as shown in Figure~4.
First, we set the incident beam spot to the center of the Si layers 
to understand the basic polarimetric response of the camera.
The simulated data were selected under the conditions described in Section\,\ref{sec: selection}.

\begin{figure}[h] 
 \begin{center}
\includegraphics[width=5in]{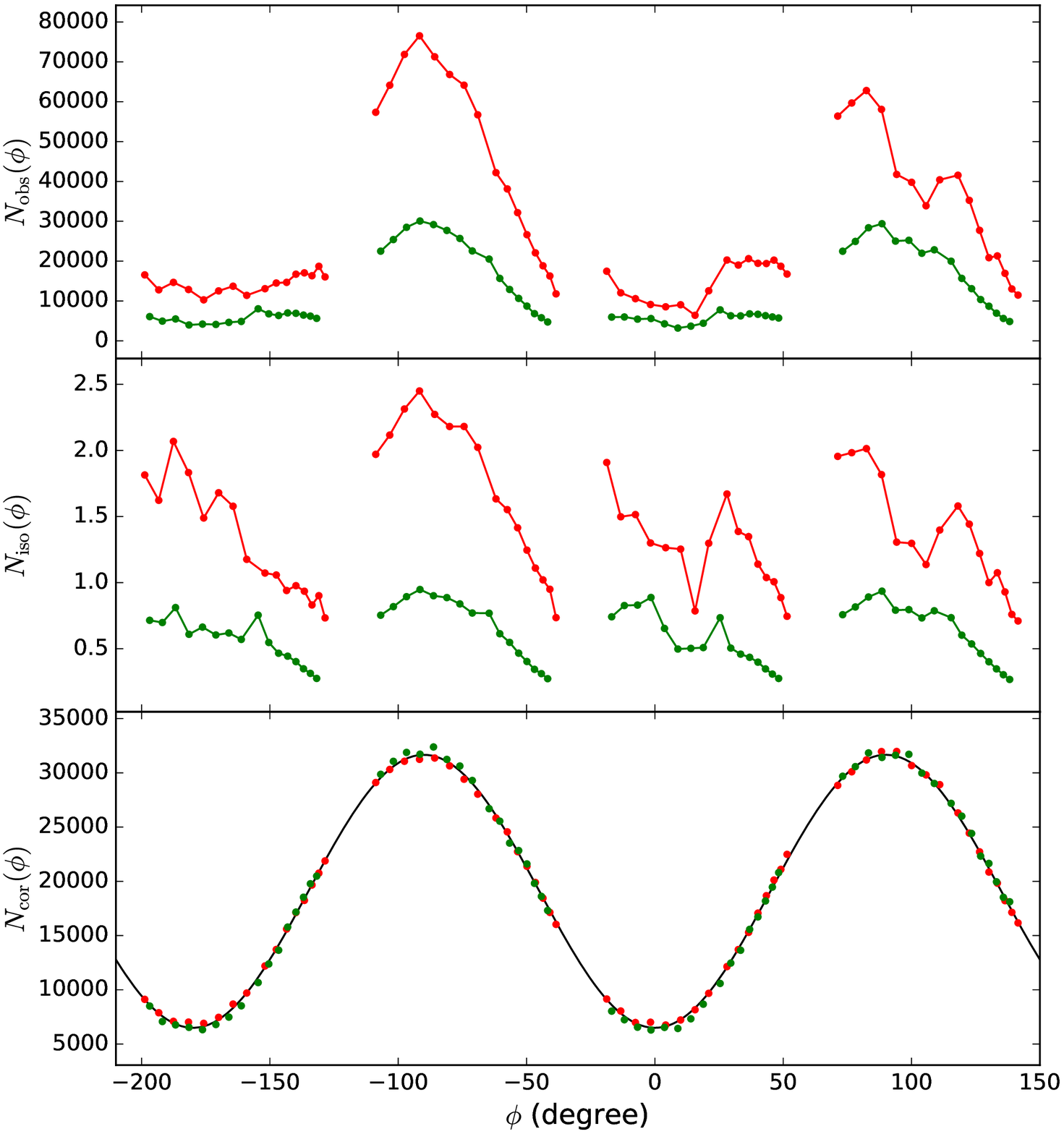}
\caption{
(Top) Azimuthal angle distribution of the simulated 100\%-polarized data
at $\pol=0^{\circ}$ for a beam energy of 122.2\,keV. 
The incident beam spot is at the center of the Si layers.
The red and green points represent data from the inner and outer layers of the side CdTe detectors, respectively.
The distribution corresponds to the $N_{\rm obs}(\phi)$ distribution in Equation~(3).
Note that the statistical errors are smaller than the marker size.
(Middle) Azimuthal angle distribution of the simulated nonpolarized beam
corresponding to the $N_{\rm iso}(\phi)$ distribution.
(Bottom) The $N_{\rm cor}(\phi)$ distribution 
obtained from $N_{\rm obs}(\phi)$ and $N_{\rm iso}(\phi)$ [see Equation~(3)].
The black line represents the best-fit model curve of Equation~(4).
\label{fig: n_corr_0}}
 \end{center}
\end{figure}

For each selected event, we calculated the azimuthal scattering angle ($\phi$)
with respect to the camera's $X$-axis using the scattered $\gamma$-ray trajectory
(see Figures~ 2 and 4).
The bin size of $\phi$ (not uniform) was $\sim 6^{\circ}$,
which corresponds to the pixel size of the CdTe sensor (3.2\,mm).
We calculated $\phi$ for the layers of the side CdTe detectors;
each side CdTe detector has 16 pixels along the $\phi$ direction so that
the total number of angle bins was $16\,{\rm pixels} \times 8\,{\rm layers} = 128$.
In this analysis, we divided the data from the eight side CdTe layers into two groups:
the data from inner four CdTe layers (hereafter the ``inner layer data") 
and the data from the outer ones (hereafter the ``outer layer data"),
as shown in Figure\,\ref{fig: setup2}.
The top panel of Figure\,\ref{fig: n_corr_0} shows the obtained azimuthal angle distribution.
This distribution corresponds to $N_{\rm obs}(\phi)$ in Section\,\ref{sec: method}.
%
The distribution of the inner layer data is systematically larger than that of the 
outer layer data,
because the scattered $\gamma$ rays first enter the inner CdTe layers 
and could be absorbed without reaching
the outer CdTe layers.
The gaps near $-120^{\circ}$, $-30^{\circ}$, and $60^{\circ}$ correspond
to the absence of the side CdTe detectors (see Section\,\ref{sec: design}).
In addition, we can see a dependency of the azimuthal scattering angle on the polarization vector,
i.e., counts near $-90^{\circ}$ and 90$^{\circ}$ are higher than those near 0$^{\circ}$ and 180$^{\circ}$.

To obtain a smooth modulation curve
that follows the sinusoidal function of Equation~(4), 
we need to correct the obtained $N_{\rm obs}(\phi)$
by the detector response for nonpolarized beam [$N_{\rm iso}(\phi)$; see Equation~(3)].
Therefore, we manufactured the same simulation data as above except with
a nonpolarized incident beam.
The azimuthal angle distribution of the simulated data is shown in the middle panel of Figure\,\ref{fig: n_corr_0}.
The distribution is bumpy, since ${\sim}10$\% of the side CdTe pixels did not work
in the 
experiment, primarily because of the bump-bonding disconnections between the sensors (Si or CdTe)
and the readout board. We note that this is taken into account in the simulation,
and the fraction of the bad pixels is much lower for the flight-model SGDs.
Using these curves,
we obtained the corrected distribution, $N_{\rm cor}(\phi)$,
as shown in the bottom panel of Figure\,\ref{fig: n_corr_0}.
The obtained $N_{\rm cor}(\phi)$ was well fit using Equation~(4),
as shown in the bottom panel of Figure~(5).
The obtained parameters were $Q = 0.66$ and $\phi_0 = 0.0^{\circ}$.
We also conducted the same analyses for the data at 194.5\,keV,
and the obtained parameters were $Q = 0.60$ and $\phi_0 = 0.1^{\circ}$.
Therefore, we confirmed the analysis procedure.

\begin{figure}[h] 
 \begin{center}
 \includegraphics[width=5in]{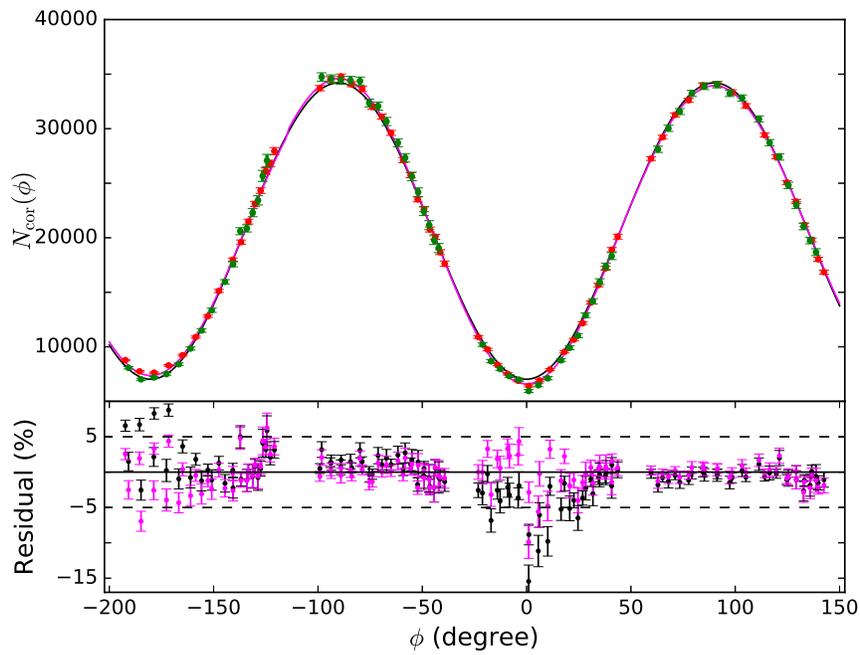}
\caption{
(Top) Same as the bottom panel of Figure\,\ref{fig: n_corr_0}
except that the incident beam spot is set to that of the experiment.
Note that most error bars are smaller than the marker size.
The black and magenta lines are the best-fit curves of
Equations~(\ref{eq: n_cor_mdl}) and (\ref{eq: n_cor_mdl_mod}) respectively.
(Bottom) The distributions of (experimental data $-$ model)$/$model.
The black and magenta points represent the residuals
when fitted by Equations~(4) and (8), respectively.
\label{fig: n_corr_22}}
 \end{center}
\end{figure}

Next, we generated the same simulation data as above at 122.2~keV,
except that the incident beam spot was set 
to that of the experiment,
which was off center by 4.8~mm along camera's $X$- and $Y$-axis (see Figure\,\ref{fig: cc_geo2}). 
Using the same analysis as above, we obtained $N_{\rm cor}(\phi)$, as shown in Figure\,\ref{fig: n_corr_22}.
The fitting with Equation~(4) resulted in $Q = 0.66$ and $\phi_0 = -0.1^{\circ}$.
Even though the obtained values were nearly consistent with those for the beam irradiated at the center
of the Si sensors,
the top and bottom panels of Figure\,\ref{fig: n_corr_22} show
that the amplitudes of the curve near $-90^{\circ}$ and $0^{\circ}$ are larger than the other peaks
($90^{\circ}$ and $180^{\circ}$, respectively).
%
This is because the differential cross section [Equation~(\ref{eq: cross})] depends not only on $\eta$
but also on $\theta$.
Strictly speaking,
Equation~(\ref{eq: n_cor_mdl}) can be applied to the distribution
where the integration of $d\sigma/d\Omega$ in the $\theta$ space is independent of $\phi$.
If this is not the case, $N_{\rm cor}(\phi)$ deviates from the model curve.
%
In this simulation, the incident beam spot is off center so that the distance
to the side CdTe detectors from the spot systematically changes in the $\phi$ direction.
This means that the viewing angle of $\theta$ changes in the $\phi$ direction.
Such a $\phi$ dependency explains why the
amplitude ($Q$) of the $N_{\rm cor}(\phi)$ distribution depends on $\phi$.
%
To compensate for this $\phi$-dependece,
we divided the distribution into two sets based on the distance to the side CdTe detectors:
``data set\,1" ($-115^{\circ} < \phi < 50^{\circ}$) and ``data set\,2" ($\phi < -115^{\circ}$ and $\phi > 50^{\circ}$).
The side CdTe detectors of the former set are more distant from the incident beam spot
than those of the latter set.
Note that we divided the inner and outer groups in the same way for simplicity.
We then modified the model curve from Equation~(\ref{eq: n_cor_mdl}) to
\begin{linenomath}
\begin{equation}
\label{eq: n_cor_mdl_mod}
N_{\rm cor}(\phi)
= \begin{cases}
A(1 + Q_{1}\cos2(\phi - \phi_0 - \pi/{2}))\ \ \ ({\rm dataset\,1}),\\
A(1 + Q_{2}\cos2(\phi - \phi_0 - \pi/{2}))\ \ \ ({\rm dataset\,2}).
\end{cases}
\end{equation}
\end{linenomath}
Different parameters were used for the amplitude ($Q_{1,2}$)
to explain the $\phi$ dependency.

The best-fit curve is shown in the top panel of Figure\,\ref{fig: n_corr_22}
(the magenta line).
The bottom panel demonstrates that the residuals decrease with respect to
the original curve.
There still remain small but statistically significant residuals (e.g., those at $\phi \sim 0^{\circ}$),
likely because the modified model function is still not perfect.
In this experiment, our goal was to evaluate the polarimetric performance of the Compton camera,
in particular, the systematic uncertainty by comparing the experimental and simulation data,
and Equation~(\ref{eq: n_cor_mdl_mod}) is sufficient for such a purpose.
The obtained parameters were $Q_1 = 0.68$, $Q_2 = 0.64$, and $\phi_0 = 0.0^{\circ}$ at 122\,keV,
while those for 193\,keV were $Q_1 = 0.64$, $Q_2 = 0.59$, and $\phi_0 = -0.3^{\circ}$.
Note that dividing the data is not necessary in orbit,
because the $\gamma$ rays enter the entire area of the Si layers and the geometrical asymmetry is canceled.

\clearpage

\subsection{Experimental Data}
\label{sec: ana_exp}
We derived the $N_{\rm cor}(\phi)$ distribution for the experimental data in a similar fashion
to that described in Section\,\ref{sec: ana_sim}
and found that the best-fit values of $\phi_0$ were systematically smaller by $\sim 1^{\circ}$
than the experimentally-set angles for all $\pol$ angles 
($0^{\circ}$, $22.5^{\circ}$, $45^{\circ}$, $90^{\circ}$, and $180^{\circ}$).
Given that the experiment setup has an uncertainty of $\sim 1^{\circ}$ 
with respect to the rotation around the beam axis,
we concluded that the discrepancies were due to misalignment.
The average of discrepancy was 1.4$^{\circ}$,
and we used this value as the offset angle around the beam axis.
Therefore, the $\pol$ angles of the experiment were
$-1.4^{\circ}$, $21.1^{\circ}$, $43.6^{\circ}$, $88.6^{\circ}$, and $178.6^{\circ}$.

\begin{figure}[h] 
 \begin{center}
 \includegraphics[width=5in]{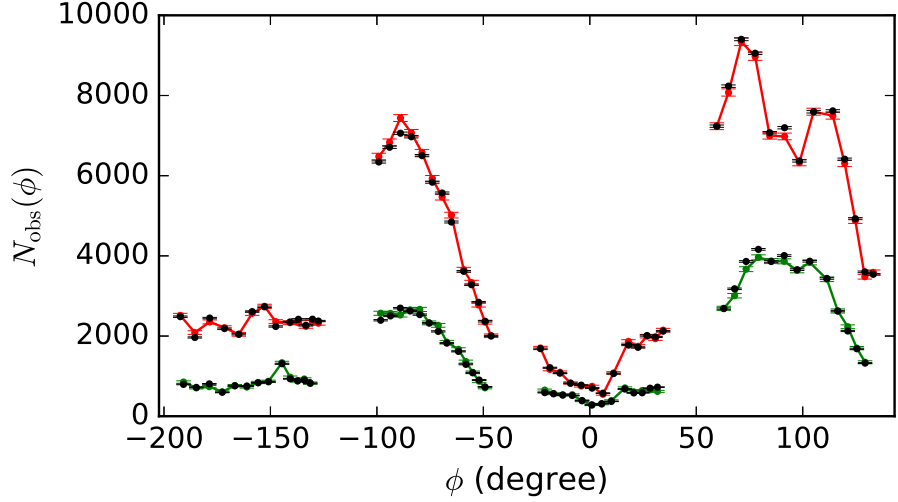}
\caption{
The $N_{\rm obs}(\phi)$ distribution at $\pol = -1.4^{\circ}$ and an energy of 122.2\,keV.
The red and green points represent the inner and outer layer data of the experiment, respectively,
while the superimposed black points show the simulated data.
The simulated data are normalized to the total angular counts of the experimental data.
%
\label{fig: n_obs_exp}}
 \end{center}
\end{figure}

We then selected the experimental data under the conditions described in Section\,\ref{sec: selection}
and derived the $N_{\rm obs}(\phi)$ distribution as shown in Figure\,\ref{fig: n_obs_exp}
for $\phi_{\rm pol}=1.4^{\circ}$.
We also superimposed the distribution of the simulation data (for $\phi_{\rm pol}=1.4^{\circ}$)
with the 100\% polarized beam in the figure.
The parameters of the simulation data were the same as those in Section\,\ref{sec: ana_sim} except that
the polar angle of the incident beam was set to 1$^{\circ}$ with respect to the normal of the Si layers 
to reproduce the misalignment between the beam and the camera in the experiment.
We also simulated data where the incident beam was exactly perpendicular to the camera;
the difference between two data sets was up to 5\% per bin in the azimuthal angle distributions,
and the discrepancy of the best-fit parameters was less than 1\%
because the difference in $N_{\rm obs}(\phi)$ was corrected by dividing by $N_{\rm iso}(\phi)$.
We unused angle bins of $-127^{\circ}\mbox{--}-121^{\circ}$, $-46^{\circ}\mbox{--}-39^{\circ}$,
$35^{\circ}\mbox{--}44^{\circ}$, and $133^{\circ}\mbox{--}143^{\circ}$,
where the trajectories of the scattered photons go through the ASICs before 
reaching the side CdTe sensors.
The mass model of the ASIC in the simulation has some uncertainty (see Section\,\ref{sec: simulation}),
and therefore, we decided not to simulate a precise response for these azimuthal angles
but instead to simply remove the data from the analysis.

We also found that the total-count ratio of the outer layer data to the inner layer data in the simulation 
was systematically larger than that in the experiment.
This is likely because the mass model of the materials between the inner and outer layer detectors 
has some uncertainties in the simulation so that the $\gamma$-rays between the detectors 
are less absorbed and/or scattered in the simulation than those in the real camera.
We compared the data for all $\pol$ and found that the total-count ratio differs systematically (by ${\sim}10\%$).
Improving the mass model is important to better reproduce the data with simulation,
and one of next steps we will persue. In this analysis, however, 
we normalized the inner and outer layer data separately in the simulation
to those in the experimental data and examined the remaining systematic uncertainty.
Figure\,\ref{fig: n_obs_exp} shows the obtained experimental and simulated $N_{\rm obs}(\phi)$ 
distributions
at $\pol = -1.4^{\circ}$ and an energy of 122.2\,keV.
Even after normalization, the figure indicates that the residuals between the 
experimental and the simulation data 
are larger than those expected from the statistical errors.
Such discrepancies are due to uncertainties in the detailed responses
of each pixel in the simulation (e.g., the energy resolution and threshold)
and should be considered as systematic uncertainties.
To evaluate the systematic uncertainty quantitatively,
we calculated the chi square value divided by the degree of freedom,
$\rchi$, between the experimental data and the simulation data
for ten data sets (the inner and the outer data at the five $\pol$ angles).
The ten calculated values range from 47.2/51 to 122.4/51, which contain
statistically unacceptable values at the 99\% confidence level.
For data sets with large $\rchi$,
we added a uniform systematic error ratio to each $\phi$ angle bin
to reduce $\rchi$ to $\simeq 1$.
The added errors were at most 3\% for ten data sets.
Using the same procedure, the systematic errors at 194.5\,keV were
evaluated to be 0\%--3\% for the ten data sets.
Therefore, the systematic errors are similar for 122.2 and 194.5\,keV.
Note that the typical statistical errors were approximately 4\% per $\phi$ angle bin.
Therefore, the simulator reproduces the shape of the experimental distribution
within a systematic uncertainty of ${\le}3\%$ except for the unused angle ranges.
We thus achieved systematic uncertainty as small as what obtained
for the SGD prototype\,\cite{Takeda10},
using the Compton camera whose design is the same as that of the flight hardware.
This uncertainty is much smaller than that of INTEGRAL/SPI in orbit evaluated in\,\cite{Dean08}.

\begin{figure}[h] 
 \begin{center}
 \includegraphics[width=5in]{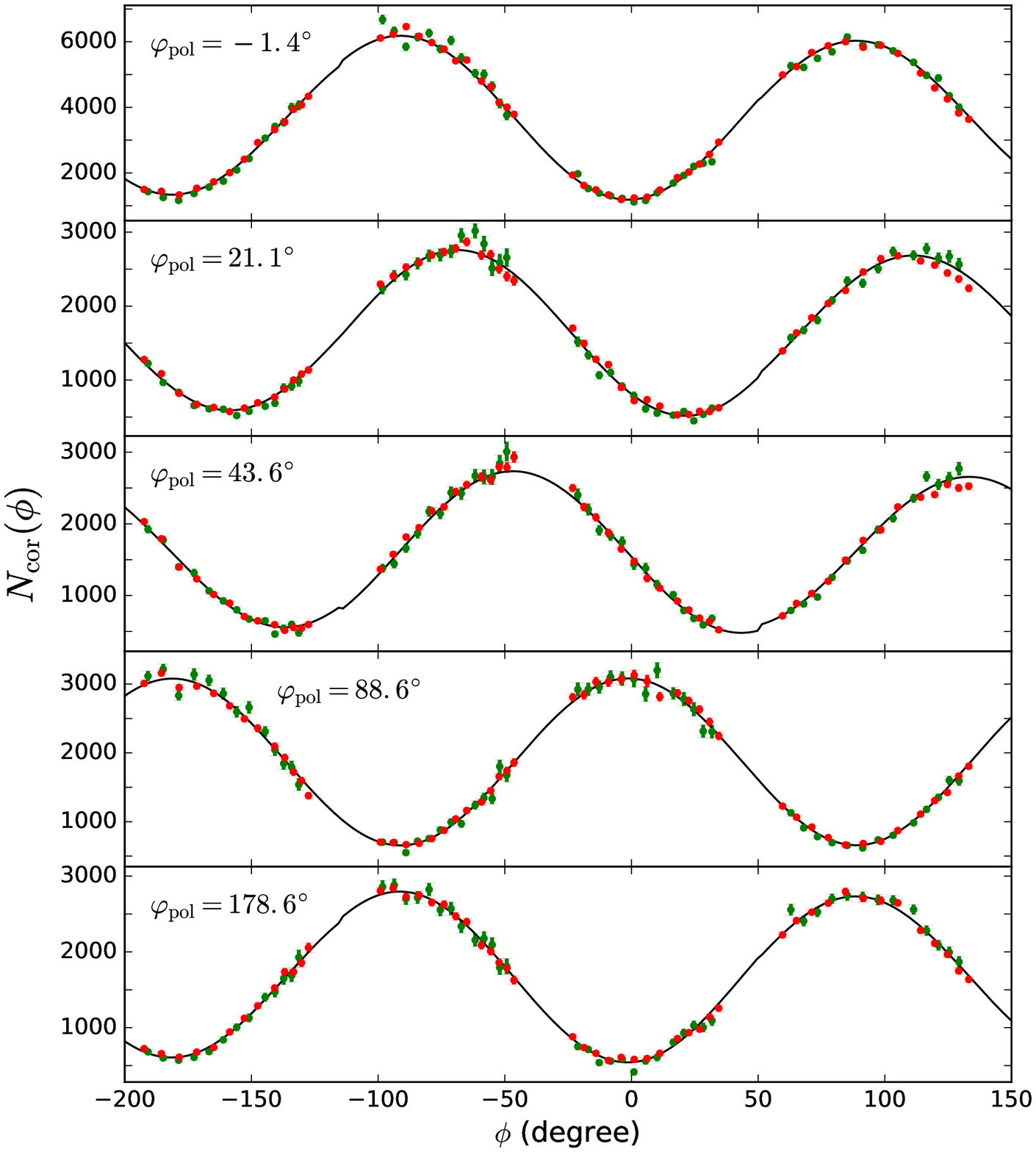}
\caption{\small
The corrected azimuth angle distributions 
for $\pol = -1.4^{\circ}$, $21.1^{\circ}$, $43.6^{\circ}$, $88.6^{\circ}$, and $178.6^{\circ}$
for a beam energy of 122.2\,keV.
The outer layer data are normalized to the average of the inner layer data.
The best-fit curves modeled by Equation~(\ref{eq: n_cor_mdl_mod}) are also plotted in black.
Note that the model curves have discontinuous values at $\phi$ of $-115^{\circ}$ and $50^{\circ}$
(see text for details).
\label{fig: n_corr_all-e1}}
 \end{center}
\end{figure}

\begin{figure}[h] 
 \begin{center}
  \includegraphics[width=5in]{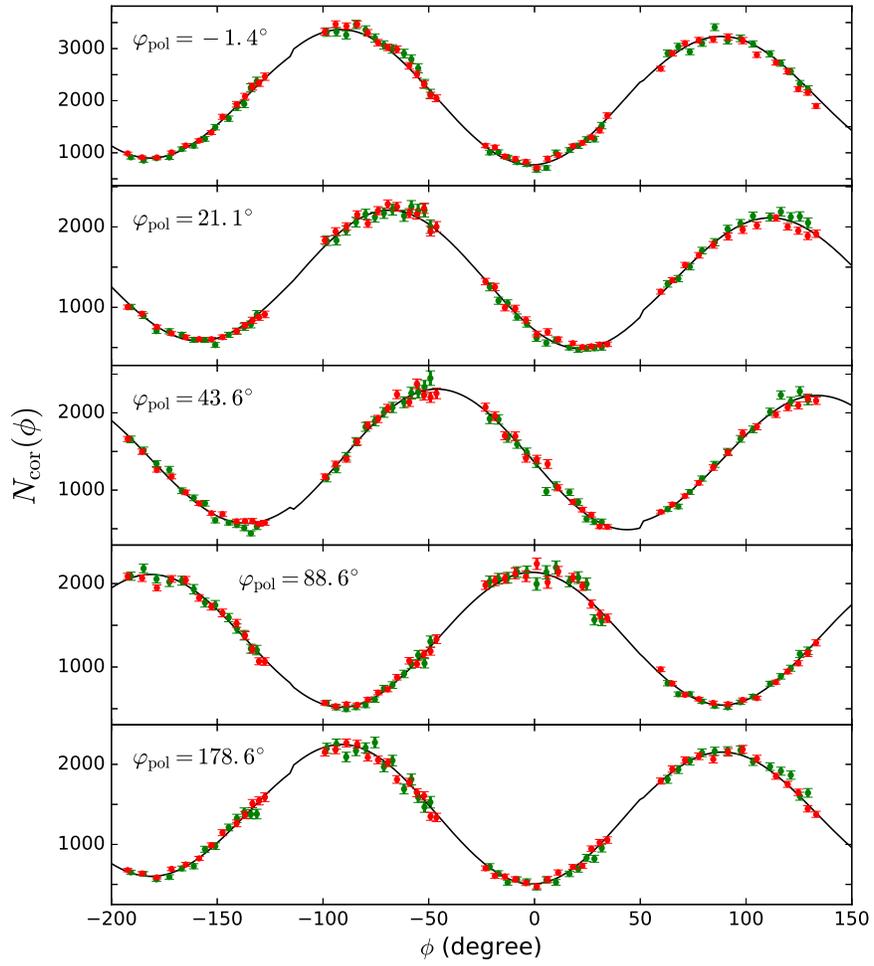}
\caption{\small
The same as Figure\,\ref{fig: n_corr_all-e1} but for a beam energy of 194.5\,keV.
\label{fig: n_corr_all-e2}}
 \end{center}
\end{figure}

We obtained the $N_{\rm cor}(\phi)$ distribution for the experimental data using the simulated $N_{\rm iso}(\phi)$
distribution for all setups and fit them by Equation~(\ref{eq: n_cor_mdl_mod}).
We used the simulated $N_{\rm iso}(\phi)$,
because it is difficult to experimentally measure the response
for nonpolarized $\gamma$-rays with the same condition.
As described above, the total count ratio of $N_{\rm obs}(\phi)$ for the 
outer layer data to the inner layer data in the simulation 
is larger than that in the experiment.
This relationship is also true for $N_{\rm iso}(\phi)$.
If we use Equation~(\ref{eq: n_cor_obs}) without corrections,
the average value of $N_{\rm cor}(\phi)$ for the outer layer data is smaller than that for the inner one.
Therefore, we normalized the average of the outer layer data to that of the inner layer data.
Figures\,\ref{fig: n_corr_all-e1} and \ref{fig: n_corr_all-e2} show the obtained
modulation curves for 122.2 and 194.5~keV, respectively.
The best-fit parameters are summarized in Table\,\ref{tbl: parameters}.
Even though the values of $\rchi$ are systematically larger than 1,
the model curves reproduce the observed data points well.
This suggests that the best-fit parameters are good indicators
of the polarimetric analyzing power of the Compton camera.

Table\,\ref{tbl: parameters} demonstrates that the measured polarization angles ($\phi_0$) are nearly
consistent with the experimentally-set values.
We calculated the differences between the measured and experimentally set angles for all $\pol$ values
and obtained a standard deviation of $0.2^{\circ}$ and $0.3^{\circ}$ for the 122.2 and the 194.5\,keV data respectively.
The uncertainties are slightly higher 
than the nominal statistical errors of ${\leq}0.2^{\circ}$.
The obtained values of $Q_1$ and $Q_2$ range 
from 0.649 to 0.701 and 0.637 to 0.653 for the inner layer data and the outer layer data, respectively, for 122.2~keV,
and
from 0.610 to 0.651 and 0.564 to 0.592 for the inner layer data and the outer layer data, respectively, for 194.5~keV.
We calculated the standard deviation between the five $\pol$ data and took it
to be the uncertainty of the $Q$ values.
Then we obtained $Q_1 =  0.677 \pm 0.018$ and $Q_2 =  0.642 \pm 0.007$ at 122.2\,keV.
The values at 194.5\,keV were calculated to be $Q_1 =  0.632 \pm 0.014$ and $Q_2 =  0.575 \pm 0.013$.
The relative standard deviations are 1.1\%--2.7\% and are larger than the nominal statistical errors of 
$\sim0.5\%$,
suggesting that the simulated response 
($N_{\rm iso}$) and/or the camera itself
has a relative systematic error of ${\le}3\%$.
We also evaluated $Q_{1}$ and $Q_{2}$ for
the simulation data
and confirmed that the simulation and experimental results were consistent within 3\%.
%
Because the incident beam was nearly 100\% polarized,
the fitted values of $Q_{1,2}$ can be considered to be experimentaly determined analyzing power ($Q_{100}$)
of the camera.
Given Equation~(\ref{eq: pol}),
an uncertainty in $Q_{100}$ indicates the potential polarimetric analyzing power of the instrument.
%
The above results show that the experimentally-determined analyzing powers are consistent with
the simulated ones within an accuracy of 3\%.

\begin{table}[h]
 \begin{center}
 \caption{Best-fit parameters of the model curve for all the setups in the experiment.}
 \begin{threeparttable}
\begin{tabular}{c|c||ccccc} \hline 
Energy & $\pol$ &  $Q_{1}$ & $Q_{2}$ & $\phi_{0}$ & $\rchi$ \\ 
${\rm [keV]}$ & ${\rm [deg]}$ & & &[deg]  & \\ \hline
122.2 &  -1.4 & $0.679 \pm 0.003$ & $0.637 \pm 0.003$ & $-1.26 \pm 0.13$ & 235.7/99 \\
      &  21.1 & $0.685 \pm 0.005$ & $0.639 \pm 0.004$ & $21.27 \pm 0.19$ & 161.4/99 \\
      &  43.6 & $0.701 \pm 0.007$ & $0.653 \pm 0.005$ & $43.30 \pm 0.16$ & 166.2/99 \\
      &  88.6 & $0.649 \pm 0.004$ & $0.648 \pm 0.004$ & $88.81 \pm 0.18$ & 132.4/99 \\
      & 178.6 & $0.675 \pm 0.005$ & $0.637 \pm 0.004$ & $178.51 \pm 0.19$ & 120.0/99 \\ \hline
194.5 &  -1.4 & $0.629 \pm 0.004$ & $0.565 \pm 0.004$ & $-1.58 \pm 0.18$ & 212.6/99 \\
      &  21.1 & $0.638 \pm 0.005$ & $0.568 \pm 0.005$ & $21.5 \pm 0.2$ & 140.2/99 \\
      &  43.6 & $0.651 \pm 0.008$ & $0.592 \pm 0.005$ & $43.76 \pm 0.18$ & 177.0/99 \\
      &  88.6 & $0.610 \pm 0.005$ & $0.590 \pm 0.004$ & $88.7 \pm 0.2$ & 147.5/99 \\
      & 178.6 & $0.633 \pm 0.005$ & $0.564 \pm 0.005$ & $178.7 \pm 0.2$ & 158.8/99 \\ \hline
 \end{tabular}
 \begin{tablenotes}
\small \item Only statistical errors are shown.
\end{tablenotes}
 \label{tbl: parameters}
 \end{threeparttable}
 \end{center}
\end{table}


\clearpage

\subsection{Polarization Sensitivity for a Celestial Object}

Finally we evaluated the polarimetric capability of the SGD on board the $\AH$ satellite 
using an MC simulation.
The minimum detectable polarization at the 99\% confidence level can be calculated as
\begin{linenomath}
\begin{equation}
 \label{eq: mdp}
{\rm MDP}_{99} = \frac{4.29}{Q_{100} \times R_S}\sqrt{\frac{R_S + R_B}{T}},
\end{equation}
\end{linenomath}
where $R_S$ and $R_B$ are the source and the background count rates respectively,
and $T$ is the observation time\,\cite{Weisskopf09}.
As a typical observation,
we considered one of the brightest X-ray sources, the Crab Nebula, whose spectrum is assumed to 
be a power law, $F(E)=K \times E^{-\Gamma}$, with a normalization of $K={\rm 11.6~photons~s^{-2}~cm^{-2}}$ at 1~keV
and a photon index of $\Gamma=2.1$, where the energy $E$ is given in keV.
Note that the Carb Nebula is so bright that the background flux is almost negligible.
In the simulation, we assumed that the incident $\gamma$ rays were perpendicular to the Si layers and
uniformly entered the entire surface of the Si layer.
The simulated $Q_{100}$ were calculated to be 0.70 and 0.61 
in the energy bands of 60--100 and 100--200\,keV, respectively.
The corresponding count rates of the SGD were expected to be 
1.3 \,counts\,s$^{-1}$ and 0.9\,counts\,s$^{-1}$, respectively.\footnote{The count rates are calculated by WebPIMMS: https://heasarc.gsfc.nasa.gov/cgi-bin/Tools/w3pimms/w3pimms.pl}
With an observation time of 100\,ks, the ${\rm MDP}_{99}$ were evaluated to be 1.7\% and 2.3\%
in the energy bands of 60--100 and 100--200\,keV, respectively.
This is much smaller than the reported polarization fraction of the Crab Nebula
(${\sim}50\%$ above ${\sim}$100~keV\,\cite{Dean08,Forot08}).
Since the relative uncertainty of $Q_{100}$ is only ${\le}3\%$ as described in Section~5.3, 
it does not affect the MDP shown above.
We note that in order to measure weak polarizations,
possible fake modulations due to the systematic uncertainty 
could affect the sensitivity
and should be investigated from ground and in-orbit data.
For less bright objects, the uncertainty in the background should also be taken into account.

\clearpage

\section{Summary}
Gamma-ray polarization is a unique probe into the geometry
of the $\gamma$-ray emission process in celestial objects.
The Soft Gamma-ray Detector onboard the X-ray observatory \textit{Hitomi}
is a Si/CdTe Compton camera and is expected to be an excellent polarimeter,
as well as a highly sensitive spectrometer because of its good angular coverage
and angular resolution for Compton scattering and its low background.
To demonstrate the polarimetric capability and to verify and calibrate
the Monte Carlo simulation program, we carried out 
a beam test of the final prototype for the SGD Compton camera at the large synchrotron facility SPring-8.
We divided the absorber into two parts, the inner and outer layers of the side CdTe sensors,
and obtained modulation factors of
0.649--0.701 (inner part) and 0.637--0.653 (outer part) at 122.2~keV and
0.610--0.651 (inner part) and 0.564--0.592 (outer part) at 194.5~keV,
at varying polarization angles with respect to the detector.
The test results suggest that the systematic uncertainty of the modulation factor is as small as ${\le}3\%$.
Using the "calibrated" simulator, we evaluated the polarimetric sensitivity of the SGD in orbit
and obtained the minimum detectable polarization (with 99\% confidence) of 1.7\% 
and 2.3\% in the energy bands of 60--100 and 100--200~keV, respectively,
for the Crab Nebula with 100~ks observation in orbit.

\section*{Acknowledgements}
This experiment was approved and conducted at the Spring-8 for proposal No. 2015B1022.
We appreciate the support of the SPring-8 staff.
This work was partially supported by 
JSPS Grant-in-Aid for Scientific Research (A) Grant No. 24244014 and
JSPS Grant-in-Aid for Scientific Research (B) Grant No. 25287059.

\clearpage

\bibliography{bib_SP8}

\end{document}